\documentclass[superscriptaddress,showpacs,amssymb,10pt,reprint,aps,prd,longbibliography,nofootinbib,floatfix]{revtex4-2}
\usepackage{graphicx,epsfig,amssymb} 
\usepackage{amsmath,amsfonts, times}
\usepackage{bm} 
\usepackage[normalem]{ulem}
\usepackage{epstopdf}
\usepackage[caption=false]{subfig}
\usepackage[usenames]{color} 
\usepackage{mathrsfs} 
\usepackage{natbib}
\usepackage{soul}
\usepackage{subfig}
\usepackage[utf8x]{inputenc}
\usepackage{tikz}
\usepackage[linktocpage,colorlinks]{hyperref}
\usepackage{cleveref}
\usetikzlibrary{decorations.pathmorphing}
\definecolor{coolblack}{rgb}{0.0, 0.18, 0.39}
\definecolor{darkred}{rgb}{0.5,0,0}
\definecolor{darkgreen}{rgb}{0,0.5,0}
\definecolor{darkblue}{rgb}{0,0,0.5}
\definecolor{lapislazuli}{rgb}{0.15, 0.38, 0.61}
\definecolor{venetianred}{rgb}{0.78, 0.03, 0.08}
\definecolor{bleudefrance}{rgb}{0.19, 0.55, 0.91}
\definecolor{dogwoodrose}{rgb}{0.84, 0.09, 0.41}
\hypersetup{colorlinks=true, linkcolor=darkgreen, citecolor=blue, urlcolor=blue}

\newcommand{\orcid}[1]{\href{https://orcid.org/#1}{\includegraphics[width=10pt]{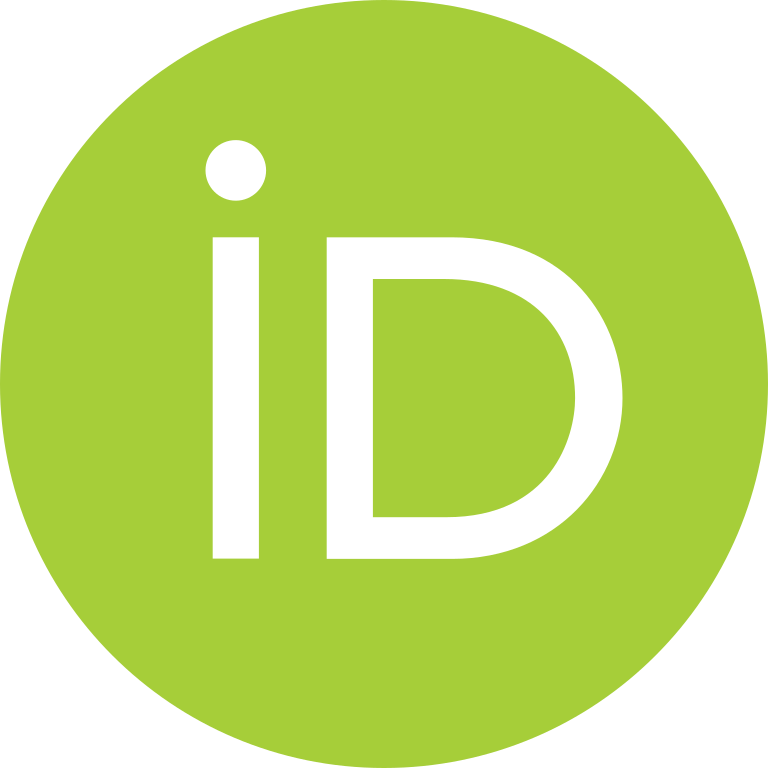}}}

\begin{document}
	\title{\large{Cosmic evolution from Lorentz-violating bumblebee dynamics and Tsallis holographic dark energy}}
    \author{E. M. Siquieri \orcid{0009-0003-8050-0370}}
    \email{eduardosiquieri@fisica.ufmt.br}
     \author{D. S. Cabral \orcid{0000-0002-7086-5582}}
	\email{danielcabral@fisica.ufmt.br}
    \author{A. F. Santos \orcid{0000-0002-2505-5273}}
    \email{alesandroferreira@fisica.ufmt.br}
	\affiliation{Programa de Pós-graduação em Física, Instituto de Física, Universidade Federal de Mato Grosso, Cuiabá, Brasil}

\begin{abstract}
In this work, the behavior, evolution, and expansion of the universe are investigated within a Lorentz-violating framework driven by Tsallis holographic dark energy. The cosmological extension is implemented through a spontaneously symmetry-breaking Bumblebee field, which is assumed to play a fundamental role in the dynamics of the universe. Estimates for key Lorentz-violating quantities are obtained, and the evolution of the Hubble parameter is analyzed from the  early universe era to the present epoch. This formulation provides an alternative perspective on the Hubble tension.
\end{abstract}
\date{\today}
\maketitle
\section{Introduction}

General Relativity (GR), proposed by Einstein in 1915, has provided the foundational description of gravitation for more than a century. The theory has been extensively tested up to the present time and remains one of the most successful frameworks in modern physics, with experimental confirmations ranging from classical tests to recent high-precision observations \cite{Dyson_IX_220_1920,Abbott_Observation_116_2016,Abbott_GW170817_119_2017,Abuter_Detection_615_2018,Akiyama_First_875_2019}. Despite this remarkable success, GR in its original form does not fully account for several key cosmological observations. In particular, the late-time accelerated expansion of the universe (associated with dark energy) and the presence and origin of dark matter remain unexplained within the standard theory \cite{Will_LivRevRel_9_2006,Riess_TheAstJour_116_1998,Perlmutter_TheAstrPhysJour_517_1999}. In addition, GR is not formulated as a quantum theory, and it is generally expected that a more fundamental description of gravity should incorporate quantum-mechanical principles.

Several theoretical frameworks have already been proposed to achieve a unification between gravitation and quantum mechanics. Among the most prominent candidates for such a fundamental description are string theory \cite{PhysToday.40.33(1987), BOOK-Springer3(1991)} and loop quantum gravity \cite{Rovelli_LivRevRel_11_2008}. However, while it is arguable whether Loop Quantum Gravity exhibits explicit Lorentz breaking symmetry, Lorentz violation effects arise in various contexts, such as string theory and effective field theory formulations \cite{Carlip_IntJouModPhysD_24_2015,PhysRevD.39.683(1989),NucPhysB.359(1991)}.

 Lorentz symmetry breaking can manifest in two ways: spontaneously and explicitly.  Spontaneous Lorentz violation manifests frequently in different states of matter \cite{nicolis}. In the context of gravity, the former involves a mechanism that chooses preferred spacetime directions and is typically associated with extremely high energy scales, although small low-energy imprints may still arise \cite{Universe.2.30(2016)}. Explicit breaking, on the other hand, is directly inserted into the theory, although current observations suggest that there is no observable explicit breaking of the Lorentz symmetry \cite{caoetal}. This possibility motivated D. Colladay and A. Kostelecký to formulate the Standard Model Extension (SME), which systematically includes in the Standard Model (SM) and GR all operators that violate Lorentz and CPT symmetries \cite{Colladay_Kostelecky_PhysRevD_55_1997,Colladay_Kostelecky_PhysRevD_58_1998}. The SME serves as a general framework for constructing effective theories with well-defined symmetry-breaking sectors.

In recent years, several theories beyond the SM and GR have been explored to identify potential signatures of high-energy unification or to establish bounds on parameters associated with symmetry breaking. These investigations span both extended quantum theories \cite{gtc1, gtc2, gtc3} and modified gravity scenarios \cite{gtc4, gtc5, gtc6}, reflecting a broad effort to understand how fundamental symmetries might be altered in more general frameworks.

A particularly relevant effective framework incorporating Lorentz violation is the bumblebee model, originally formulated with structural similarities to electromagnetism \cite{PhysRevD.40(1989)}. In this theory, a vector field--referred to as the bumblebee field--develops a nonvanishing vacuum expectation value (VEV), thereby inducing a spontaneous breaking of Lorentz symmetry. The resulting symmetry breaking modifies the Einstein field equations and gives rise to potentially detectable imprints in gravitational phenomena. Extensive studies of the bumblebee framework and its implications can be found in Refs.~\cite{PhysRevD.69.105009(2004),Capelo_PhysRevD_91_2015,Santos_ModPhysLettA_30_2015,Jesus_ModPhysLettA_34_2019,Jesus_IntJourModPhysA_35_2020}.

On the other hand, the observational confirmation of the universe’s accelerated expansion \cite{Riess_TheAstJour_116_1998,Perlmutter_TheAstrPhysJour_517_1999} initiated an extensive line of research aimed at identifying the underlying mechanism responsible for this phenomenon. Both modifications to GR and the introduction of dark energy components emerged as leading hypotheses, giving rise to a wide range of theoretical models \cite{Caldwell_PhysRevLett_80_1998,Gao_PhysRevD_79_2009,Felice_LivRevRel_13_2010}.

Within this broader context of cosmic acceleration, one of the most pressing open problems in modern cosmology is the so-called Hubble tension. This issue stems from the persistent discrepancy between the values of the Hubble constant inferred from early-universe (indirect) observations and those obtained through late-universe (direct) measurements \cite{hubbletension}. Motivated by this inconsistency, numerous studies have explored generalized and modified gravitational frameworks as possible resolutions \cite{hubblemod1, hubblemod2, hubblemod3}.

In parallel, foundational developments in black hole thermodynamics by Bekenstein and Hawking \cite{Bekenstein.PhysRevD.7(1973),Hawking.CommMathPhys.31(1973),Hawking.Nature.248(1974)} played a crucial role in the establishment of the holographic principle \cite{tHooft:1993itb}. One approach to explaining cosmic acceleration draws on this principle by applying holographic considerations to the dark energy sector, leading to the holographic dark energy (HDE) paradigm, in which the number of degrees of freedom of the system is effectively reduced.

In this context, the generalization of entropy proposed by Tsallis \cite{Tsallis.JornStatPhys.52(1988)}—which has played an increasingly relevant role in gravitational and cosmological applications within statistical mechanics—offers a natural framework for extending the holographic dark energy paradigm. In particular, Tsallis holographic dark energy (THDE) incorporates corrections to the usual area-dependent entropy relations \cite{Tavayef_PhysLettB_781_2018}, including those associated with black hole thermodynamics \cite{Tsalli.EurPhysJC.73(2013)}. The resulting energy density depends on the nonextensive parameter $\delta$, which determines distinct classes of dark energy models. Tsallis originally argued that $\delta = 3/2$ provides the appropriate correction to restore extensivity in black hole entropy \cite{Tsalli.EurPhysJC.73(2013)}, and recent analyses suggest that observationally favored values indeed lie close to this range \cite{Tsallis.PhysLettB.861(2025)}. Other choices of the holographic parameter remain of interest as well; for instance, $\delta = 1$ reproduces the standard Bekenstein–Hawking entropy and recovers the conventional HDE formulation \cite{Tavayef_PhysLettB_781_2018, Wang.PhysRep.696(2017)}.

With these considerations in mind, this work explores the cosmic evolution resulting from the spontaneous Lorentz-symmetry breaking induced by the Bumblebee field in the presence of Tsallis holographic dark energy, offering new insights into the Hubble tension.

The paper is organized as follows. Section \ref{secframework} introduces the theoretical foundation of the analysis, including the Bumblebee model in Subsec.~\ref{secbumblebee}, the cosmological background in Subsec.~\ref{secmetric}, and the discussion of the Hubble tension in 
Subsec.~\ref{sectension}. Subsection \ref{seccontent} describes the matter content adopted in this study, while Subsec.~\ref{secpotential} presents the mechanism responsible for the spontaneous Lorentz symmetry breaking. Section \ref{secresults} contains the results obtained for three different choices of the holographic parameter: $\delta = 2$ (Subsec.~\ref{secresults2}), $\delta = 3/2$ (Subsec.~\ref{secresults3/2}), and $\delta = 1$ (Subsec.~\ref{secresults1}). The validity and stability of the presented results, particularly regarding the Hubble tension correction, are addressed in Subsec.~\ref{secresultsvalidty}. Finally, concluding remarks are given in Sec.~\ref{secconclusion}. 

Throughout this paper, natural units are not employed,  instead, we will use the International System of Units (SI). This choice simplifies the interpretation of the results, enables direct proposals for the free parameters, and facilitates comparison with observationally motivated quantities.

\section{Theoretical framework}
\label{secframework}

The main goal here is to derive the equations of motion that govern the evolution of a universe containing a Lorentz-violating Bumblebee field and dark energy modeled through the Tsallis holographic principle. To this end, we begin by presenting the theoretical framework of spontaneous symmetry breaking, along with the necessary formalism, mechanisms, and results that enable us to describe the cosmic expansion from the early universe to the present epoch.

\subsection{Bumblebee Model}
\label{secbumblebee}

In this subsection, the Bumblebee model is introduced. Within the framework of General Relativity, a Lorentz-violating theory was proposed by Kosteleck\'{y} and Samuel~\cite{PhysRevD.40(1989)}, inspired by the structure of quantum electrodynamics (QED). In this model, a dynamical vector field $B_\mu$ acquires a nonvanishing vacuum expectation value (VEV),
\begin{eqnarray}
\langle B_{\mu} \rangle = b_{\mu} \neq 0,
\end{eqnarray}
which leads to the spontaneous breaking of Lorentz symmetry and selects a preferred direction in spacetime~\cite{PhysRevD.69.105009(2004), PhysRevD.77.065020(2008)}.

With this feature, modifications to General Relativity can naturally be introduced. The resulting framework, known as the Bumblebee model, admits a variety of interaction terms -- both minimal and nonminimal -- rendering it a rich and versatile theoretical structure. A general Lagrangian describing the Bumblebee model can be expressed as
 \begin{align} {\label{eq:LagrangianaGeralBumblebee}}
    \mathcal{L}_{B}
    &=  
    \frac{1}{16\pi G}  (R - 2\Lambda) 
    + 
    \sigma_{1}  B^{\mu} B^{\nu} R_{\mu\nu} 
    +  
    \sigma_{2}  B^{\mu} B_{\mu}R \nonumber \\
    & 
    - 
    \frac{1}{4} \tau_{1}  B^{\mu\nu} B_{\mu\nu} 
    + 
    \frac{1}{2} \tau_{3}  \nabla_{\mu} B^{\mu} \nabla_{\nu}B^{\nu}
    + 
    \frac{1}{2} \tau_{2}  \nabla_{\mu} B_{\nu} \nabla^{\mu}B^{\nu}\nonumber\\
    &
    -
    V+\mathcal{L}_M
     \,,
\end{align}
where $\sigma_i$ and $\tau_j$ are coupling parameters that may or may not be mutually dependent~\cite{PhysRevD.77.065020(2008)}. The potential $V = V(B^{\mu} B_{\mu} \pm b^{2})$ is defined to break Lorentz invariance by inducing a nonzero vacuum expectation value $b$, thereby selecting a preferred direction in spacetime. The Bumblebee field strength $B_{\mu\nu}$ is defined as
\begin{eqnarray}{\label{eq:Definicao Intensidade de B}}
B_{\mu\nu} \equiv \partial_{\mu} B_{\nu} - \partial_{\nu} B_{\mu} ,
\end{eqnarray}
in direct analogy with the electromagnetic field tensor~\cite{PhysRevD.40(1989), PhysRevD.69.105009(2004), PhysRevD.77.065020(2008)}.  The operator $\nabla_\mu$ is the gravitational covariant derivative. Finally, $\mathcal{L}_M$ denotes the matter Lagrangian density.

The elegance of this formulation lies in its direct correspondence with the Lagrangian of the gravitational sector of the Standard-Model Extension (SME),
{\small
\begin{equation}
\mathcal{L}{\text{g}}^{\text{SME}}
= \frac{1}{16\pi G} \left[ (1 - u) R - 2\Lambda + s^{\mu\nu} R_{\mu\nu} + t^{\kappa\lambda\mu\nu} R_{\kappa\lambda\mu\nu} \right],
\label{eq06}
\end{equation}
}
from which the following identifications can be made:
\begin{eqnarray}
&&\frac{u}{16\pi G} = -\sigma_2 \xi B^{\alpha} B^{\beta} g_{\alpha\beta},
\quad
\frac{s^{\mu\nu}}{16\pi G} = \sigma_1 B^{\mu} B^{\nu},\nonumber\\
&&\nonumber\\&&
t^{\kappa\lambda\mu\nu} = 0.
\end{eqnarray}
In other words, the components $u$ and $s^{\mu\nu}$ of the Lorentz-violating SME gravitational sector correspond to the respective terms in the Bumblebee theory presented in \eqref{eq:LagrangianaGeralBumblebee} \cite{PhysRevD.69.105009(2004),PhyRevD.74.045001(2006)}.

Since the most general form of the Lagrangian in~\eqref{eq:LagrangianaGeralBumblebee} contains several coupling terms, we restrict our analysis to a simplified case defined by $\sigma_{2} = \tau_{2} = \tau_{3} = 0$, $\tau_{1} = 1$, and $\sigma_{1} = \xi / 2k$.  The reason for this choice is to maintain the standard Maxwellian form of the kinetic term and address only the interaction term related to the $s_{\mu\nu}$ Lorentz-violating SME parameter, as it describes more elegant and interesting physics. Under these conditions, the Lagrangian takes the form
{\small
\begin{eqnarray}{\label{eq:LagrangianaAcoplando.B.R}}
\mathcal{L}_{B\xi} = \frac{1}{2k}
\left(
R - 2\Lambda + \xi B^{\mu} B^{\nu} R_{\mu\nu}
\right)
- \frac{1}{4} B^{\mu\nu} B_{\mu\nu}
- V + \mathcal{L}_M,
\end{eqnarray}
}
where $\xi$ denotes the coupling constant and $k = 8\pi G / c^3$ is the gravitational constant. It is important to remark that, although physically motivated by its correspondence with the SME and the standard Bumblebee model, this particular choice of coupling constants—when combined with the spontaneous symmetry breaking mechanism—does not automatically guarantee the absence of instabilities, such as ghost modes or other pathologies. Accordingly, we treat the model as an effective description, assuming that any potential unphysical effects remain under control within the phenomenological regime of interest, as discussed throughout the paper.

By setting $\Lambda = 0$ in this framework, the equation of motion for the Bumblebee field is obtained as
\begin{eqnarray}{\label{eq:Dinamica do Campo B com Acoplamento}}
\nabla_{\alpha} B^{\alpha\beta}
= 2 \left( V' B^{\beta} - \frac{\xi}{2k} B_{\alpha} R^{\alpha\beta} \right),
\end{eqnarray}
while the corresponding gravitational field equation takes the form
\begin{eqnarray}
G_{\alpha\beta}
    &=&
    k T_{\alpha\beta} 
    + k \Bigl[
            2 V' B_{\alpha}B_{\beta}  
            + {B_{\beta}}^{\nu}B_{\alpha\nu} \nonumber\\ 
            &&- \left(
                V  
                + \frac{1}{4} B^{\mu\nu} B_{\mu\nu} 
            \right)g_{\alpha\beta}
        \Bigl]+\xi \Bigl\{  
             \frac{1}{2}B^{\mu}B^{\nu}R_{\mu\nu} g_{\alpha\beta}\nonumber\\
            &&- B_{\alpha}B^{\nu} R_{\nu\beta} - B_{\beta}B^{\mu} R_{\mu\alpha} + \Bigl[
                 \frac{1}{2} \nabla_{\mu} \nabla_{\alpha} \left( B^{\mu}B_{\beta} \right)\nonumber\\
                 &&+ \frac{1}{2} \nabla_{\nu} \nabla_{\beta} \left( B^{\nu}B_{\alpha} \right)  
                - \frac{1}{2} \nabla_{\mu} \nabla_{\nu} \left( B^{\mu}B^{\nu}\right) g_{\alpha\beta} \nonumber\\
                &&- \frac{1}{2} \square \left(B_{\alpha}B_{\beta} \right)\Bigl]
        \Bigl\}.\label{eq:RG Modificada com Acoplamento}
\end{eqnarray}
Since $G_{\alpha\beta}$ denotes the Einstein tensor, Eq.~\eqref{eq:RG Modificada com Acoplamento} corresponds to the Einstein field equations modified by the presence of the Lorentz-violating Bumblebee field, which couples to gravity with strength $\xi$ through the spacetime curvature.

To carry out a cosmological analysis of this model, both field equations must be solved within a specified background geometry, together with an appropriate choice of matter content.

\subsection{Background metric}
\label{secmetric}

To analyze the field equations, the flat Friedmann-Lema\^{i}tre-Robertson-Walker (FLRW) metric is adopted. As a first step, some relevant cosmological scenarios are examined. It is worth noting that the following discussion applies to the de Sitter case. The metric describing this spacetime is given by
\begin{eqnarray}
    \mathrm{d}s^{2} = -\mathrm{d}t^{2} + a(t)^{2}\left( \mathrm{d}r^{2} + r^{2}\mathrm{d}\theta^{2} + r^{2}\sin^{2}{\theta}\mathrm{d}\phi^{2} \right),\label{eq05}
\end{eqnarray}
where $a(t)$ is the scale factor. By substituting this metric into the standard framework of GR, one obtains the Friedmann equations, which yield the following solution for the scale factor
\begin{equation}
    a(t)=\exp(H_0(t-t_0)).\label{eq10}
\end{equation}
This solution describes an accelerated expansion of the Universe. The parameter $H_0$ denotes the Hubble constant, whose value, as determined by direct measurements, is approximately $72.3~\text{km\,s}^{-1}\text{Mpc}^{-1}$ according to Ref.~\cite{hubblevalue1}, and $69.8~\text{km\,s}^{-1}\text{Mpc}^{-1}$ as reported in Ref.~\cite{hubblevalue2}. In the following analysis, we adopt the former value for consistency.

The constant $a(0)$ represents the normalization determined by the initial conditions. This normalization is chosen such that $a(t_0) = 1$ at the present epoch, where $t_0 = 13.8~\text{Gyr}$ denotes the age of the Universe.

An essential quantity for characterizing the cosmic expansion in this framework is the deceleration parameter, defined as
\begin{equation}
q = -\frac{\ddot{a}a}{\dot{a}^{2}},
\end{equation}
which measures the rate at which the cosmic expansion slows down or accelerates. For the cosmological dynamics described by Eq.~\eqref{eq10}, one finds $q = -1$, corresponding to a constant accelerated expansion of the Universe. 

When examining the Hubble constant $H_0$, which characterizes the current expansion rate of the Universe as given by Eq.~\eqref{eq10}, one encounters certain challenges, most notably the so-called Hubble tension. This discrepancy will be discussed in more detail in the following section.

\subsection{Hubble Tension}
\label{sectension}

The Hubble tension arises from the existence of two independent approaches to determine the value of the Hubble constant. In the first method, $H_0$ is inferred from early-universe observations, where data from the cosmic microwave background (CMB) are collected and extrapolated to the present epoch within the $\Lambda$CDM framework~\cite{hubbletension}. One of the most precise determinations obtained through this approach yields $H_0 = 67.4~\text{km\,s}^{-1}\text{Mpc}^{-1}$~\cite{planck2028}. In contrast, the late-universe measurements, obtained empirically from local observations -- such as Cepheid-calibrated Type Ia supernovae and complementary distance indicators -- provide a higher value, $H_0 = 73.2~\text{km\,s}^{-1}\text{Mpc}^{-1}$~\cite{riess}.

This discrepancy between early- and late-time determinations defines what is known as the Hubble tension. In what follows, attention is focused on the first type of measurement, which is based on observations of the early universe.

The inflationary era represents the first dynamical phase of the universe, occurring immediately after its origin. It is supported by both observational evidence and theoretical predictions from modern cosmology. This period provides a natural explanation for the initial conditions that governed the subsequent cosmic evolution.

From the Planck measurements of CMB anisotropies, an upper limit of $r_{0.002} < 0.10$ was obtained at a 95\% confidence level for the tensor-to-scalar ratio, a parameter that quantifies the relative amplitude of primordial gravitational waves~\cite{rmeasurement}. This bound, in turn, imposes the following constraint on the Hubble parameter during the inflationary era
\begin{eqnarray}
\frac{H_\text{inf}}{M_P} < 2.5 \times 10^{-5},
\end{eqnarray}
expressed in natural units. Here $M_P$ denotes the Planck mass, and is given by the relation $M_P^2=\hbar c/G$, in SI units. In a more conventional form, this corresponds to
\begin{eqnarray}
H_\text{inf} < 1.43 \times 10^{58}\, \text{km}\,\text{s}^{-1}\,\text{Mpc}^{-1}.\label{eq14}
\end{eqnarray}

In the present analysis, this upper bound is adopted as the initial condition for the dynamical system under consideration.  In addition, this limit can be used to establish the connection between direct and indirect measurements of this quantity.

 Before proceeding, however, it is necessary to specify the matter content responsible for driving the universe's expansion.

\subsection{Matter and energy content of the Universe}
\label{seccontent}

With these considerations in place, the dynamics of the universe can now be derived within the Lorentz-violating framework, taking into account the contribution of matter. This formulation allows the model to be applied to a cosmological scenario that incorporates dark energy, offering a possible mechanism to account for the accelerated expansion of the universe. In the present analysis, only a holographic dark energy component is considered as the matter content.

Li proposed the expression $\rho_\Lambda = 3 \zeta^2 M_P^2 L^{-2}$ for the holographic energy density (HED)~\cite{Li_PhysLettB_603_2004}, where $\zeta$ is a dimensionless constant and $L$ represents a characteristic infrared (IR) length scale. This formulation is now widely adopted as the standard definition of the holographic energy density. Li's proposal is motivated by the earlier work of Cohen et al., who established a connection between ultraviolet (UV) and infrared (IR) cutoffs in quantum field theory to address the inconsistencies arising from the non-extensive nature of Bekenstein entropy. The corresponding relation is given by $L^3 \Lambda^4 \lesssim L M_P^2$, where $\Lambda$ represents the UV cutoff~\cite{Cohen.PhysRevLett.82(1999)}.

Taking into account the Tsallis generalization of entropy, which modifies the standard Boltzmann-Gibbs framework to restore the extensivity of black hole entropy~\cite{Tsalli.EurPhysJC.73(2013)}, Tavayef et al. proposed that the Tsallis entropy could serve as a new constraint on the UV and IR cutoffs originally introduced by Cohen et al.~\cite{Tavayef_PhysLettB_781_2018}.

In this formalism, the Tsallis entropy is expressed as
\begin{eqnarray}
S_\delta = \varphi A^\delta,
\end{eqnarray}
where $\varphi$ is a proportionality constant and $\delta$ is the nonextensive parameter. The standard Bekenstein-Hawking entropy is recovered in the limit $\delta = 1$, with $\varphi = k_{B} c^{3} / (4 G \hbar)$. When natural units are adopted ($k_B = c = \hbar = 1$), this expression simplifies to $\varphi = 1 / (4G)$~\cite{Bekenstein.PhysRevD.7(1973),hawking2}.

This reasoning led Tavayef et al. to propose the expression
\begin{eqnarray}
\rho_{\text{THDE}} = \mathcal{B}_\delta L^{2\delta - 4},
\end{eqnarray}
for the Tsallis holographic dark energy (THDE) density~\cite{Tavayef_PhysLettB_781_2018}, where $\mathcal{B}_\delta$ is a model-dependent constant that varies with the nonextensive parameter $\delta$. In this framework, the Hubble horizon is adopted as the infrared (IR) cutoff, such that $L = H^{-1}$, with $H \equiv \dot{a}/a$ denoting the Hubble parameter.

In this framework, the THDE density is defined as
\begin{eqnarray}
\rho = \frac{Q_\delta \hbar}{c^{3 - 2\delta}} \left( \frac{\pi c^{3}}{G \hbar} \right)^{\delta} \left( \frac{\dot{a}}{a} \right)^{4 - 2\delta}, \label{eq04}
\end{eqnarray}
where $Q_{\delta}=\left(\frac{m}{M_p}\right)^{2\delta}$is a dimensionless constant, and $\delta$ denotes the holographic (nonextensive) parameter, with $m$ representing a characteristic mass scale. It is worth noting that, for $\delta=1$, one recovers, in natural units,
\begin{eqnarray}
\rho \propto M_P^2 \left( \frac{\dot{a}}{a} \right)^2,
\end{eqnarray}
which is consistent with the standard expression $\rho_D=3\zeta^2M_p^2/L^2$ introduced in Ref.~\cite{Li_PhysLettB_603_2004}.

Accordingly, the energy-momentum tensor can be written as
\begin{eqnarray}
T_{\mu\nu} = \rho U_\mu U_\nu,
\end{eqnarray}
where $\rho$ is given by Eq.~\eqref{eq04} and $U^\mu=(c,0,0,0)$ is the four-velocity of a comoving frame. Substituting this expression into Eq.~\eqref{eq:RG Modificada com Acoplamento} allows one to fully characterize the dynamics of the model.

Finally, it remains to discuss the form of the potential, which plays a fundamental role in introducing the Lorentz-violating character of the theory.

\subsection{The potential}
\label{secpotential}

In this subsection, the Lorentz-violating potential is specified. A natural and straightforward choice for this potential is a Higgs-like self-interaction~\cite{higgs1, higgs2}, given by
\begin{eqnarray}
    V(X)=\frac{1}{2}\lambda X^2,\label{eq02}
\end{eqnarray}
as proposed by \cite{PhysRevD.69.105009(2004)}, where $\lambda$ is a coupling constant and  
\begin{eqnarray}
    X=g^{\mu\nu}B_{\mu}B_\nu\pm b^2,\label{eq03}
\end{eqnarray}
acts as the source of the mechanism responsible for generating the vacuum expectation value (VEV) $b$, as discussed in~\cite{PhysRevD.69.105009(2004)}. The sign depends on the metric signature, being positive (negative) when the temporal component is negative (positive). 

From Eqs.~\eqref{eq02} and~\eqref{eq03}, the following contribution can be extracted from the Lagrangian~\eqref{eq:LagrangianaAcoplando.B.R}
\begin{align}
    \mathcal{L}_{B\xi}&=\frac{1}{2k} 
    \left(
        R + \xi B^{\mu}B^{\nu}R_{\mu\nu}
    \right)+\mathcal{L}_M\nonumber\\
    &-\frac{1}{4} B^{\mu\nu}B_{\mu\nu} 
    -\frac{1}{2}\lambda (b^4+B^4)+\lambda b^2B^2.\label{eq23}
\end{align}
By dimensional analysis, the coupling constant $\xi$ has the same dimension as $k$, whereas $B^2$ carries its inverse dimension. Furthermore, $\lambda$ has units of $\hbar^{-1}$.

By comparing Eq.~\eqref{eq23} with the Proca theory~\cite{proca}, the mass of the Bumblebee field can be identified as
\begin{eqnarray}
m_B=\frac{\sqrt{2\lambda}\,|b|\,\hbar}{c},\label{eq11}
\end{eqnarray}
which can be used to determine either the mass $m_B$ or the self-interaction parameter $\lambda$. However, to assign numerical values or establish upper bounds for these free parameters, additional phenomenological or observational comparisons are required.

When the Bumblebee model~\eqref{eq:LagrangianaAcoplando.B.R} is compared with the gravity sector of the SME~\eqref{eq06}, and only the timelike component is considered, an upper bound of the form
\begin{eqnarray}
\xi b^{\mu}b^{\nu}=s^{\mu\nu}<10^{-11}\label{eq07}
\end{eqnarray}
can be established, based on the constraints reported in~\cite{kosteleckydatatable}. At first glance, this bound does not directly determine the indivi\-dual values of the Bumblebee vacuum expectation value (VEV) or its coupling to gravity, leaving some freedom in the choice of these parameters. Consequently, two possible scenarios arise: either $|b^\mu|$ is very small, allowing for relatively large values of the coupling constant $\xi$, or the gravitational coupling is weak while the Bumblebee field attains appreciable magnitudes.

In this work, the latter scenario is adopted, in which the vacuum expectation value (VEV) is of order unity ($\sim 1$), and the coupling constant $\xi$ is assumed to be very small. This choice is consistent with the fact that gravity is a weak interaction compared to the other fundamental forces. As a consequence, the experimental detection of any Bumblebee excitation remains beyond current experimental reach. Furthermore, this assumption is supported by the notion that, at the Planck scale, the Lorentz-violating field may have played a significant role in the early evolution of the universe, whereas its effects are negligible in the present epoch.

With these considerations in place, the framework is now complete for evaluating key cosmological quantities, such as the expansion rate of the universe, the Hubble parameter, and the deceleration parameter.

\section{Results}
\label{secresults}

With the formal framework established in the previous sections, the dynamics of the universe is now investigated in the presence of the Bumblebee field. To preserve spatial isotropy, the field is assumed to possess only a temporal component,
\begin{equation}
B_\mu = \big(B(t), 0, 0, 0\big),\label{eq27}
\end{equation}
so that compatibility with the homogeneous and isotropic FLRW background is maintained. In this configuration, Lorentz symmetry is broken solely along the temporal direction, while spatial isotropy remains preserved.

Based on this framework and following Eq.~\eqref{eq:Dinamica do Campo B com Acoplamento}, the following relation is obtained:
\begin{equation}
\left(V^\prime - \frac{3\xi}{2k}\frac{\ddot{a}}{a}\right)B = 0. \label{eq08}
\end{equation}
Assuming, without loss of generality, that the Bumblebee field vanishes only asymptotically and adopting the potential defined in Eqs.~\eqref{eq02} and \eqref{eq03}, Eq.~\eqref{eq08} can be expressed as
\begin{equation}
B^2 = b^2 - \frac{3\xi}{2\lambda k c^2}\frac{\ddot{a}}{a}. \label{eq15}
\end{equation}

On the other hand, by inspecting the Einstein equations in Eq.~\eqref{eq:RG Modificada com Acoplamento}, together with the holographic dark energy density given in Eq.~\eqref{eq04} and the FLRW metric in Eq.~\eqref{eq05}, the following modified Friedmann equation is obtained
\begin{align}
\left( \frac{\dot{a}}{a} \right)^{2} \left( 1 - \xi B^{2} \right)
&= \frac{\lambda kc^2}{6}\left(B^2-b^2\right)^2 + \xi \left( \frac{\dot{a}}{a} \right) B \dot{B}\nonumber\\&+\frac{kQ_\delta\hbar}{3c^{2-2\delta}}\left( \frac{\pi c^{3}}{G\hbar}\right)^{\delta}\left(\frac{\dot{a}}{a}\right)^{4-2\delta}.\label{eq09}
\end{align}
This coupled system of differential equations does not admit an analytical solution; however, it can be investigated within specific temporal regimes or by applying suitable approximations.

The two preceding equations of motion describe the universe's evolution under the influence of dark energy and a Lorentz-violating massive photon-like field. The analysis is carried out for three representative cases:  $\delta = 2$, $\delta = 3/2$, and $\delta = 1$. These values are well motivated: $\delta = 1$ yields the Bekenstein-
Hawking entropy, $\delta = 2$ reproduces $\Lambda$CDM model, and $\delta = 3/2$ maintains thermodynamic extensivity consistent with observations \cite{Tsallis.PhysLettB.861(2025)}.

\subsection{Solutions for \texorpdfstring{$\delta=2$}{delta = 2}}
\label{secresults2}

In this subsection, we first analyze the asymptotic late-time limit of the model, which corresponds to the case $\delta = 2$. For this particular choice, and assuming the system has relaxed to the configuration where Lorentz-violating effects are negligible ($\xi \to 0$), the modified Friedmann equation takes the following form when written in terms of the fundamental constants
\begin{equation}
    \left( \frac{\dot{a}}{a} \right)^{2}=\frac{8\pi^3c^5m^4}{3\hbar GM_p^4}\equiv H_0^2,\label{eq18}
\end{equation}
which corresponds to the standard GR-like solution with a cosmological constant $\Lambda$, where the THDE energy density $\rho$ remains constant. It is worth noting that this regime represents the equilibrium state reached at the end of the cosmological evolution, where $B = b$ everywhere, representing the true vacuum state of the theory.

This equation allows us to establish the boundary conditions for the Hubble constant and the mass $m$ at the present epoch, with $H_0 = 72.3$ km s$^{-1}$ Mpc$^{-1}$ adopted as the present value. Accordingly, one obtains
\begin{equation}
m = 2.55\times10^{-3}\,\text{eV},\label{eq12}
\end{equation}
which fixes the value of the free parameter $m$ consistent with current observations. In this attractor regime, the Bumblebee field exhibits no evolution. Henceforth, this mass parameter $m$ will be identified with the Bumblebee mass $m_B$ defined in Eq.~\eqref{eq11}.

To investigate the role of spontaneous Lorentz-symmetry breaking, the behavior of the parameters $\xi$ and $\lambda$ must be exami\-ned.  The observational constraint presented in Eq.~\eqref{eq07} imposes a limit on the product $\xi b^{2}$. Since the model possesses freedom in defining the individual values of these parameters, we fix the energy scale of the background field to unity, i.e., $b^{2} = 1\,\text{J\,s\,m}^{-2}$, to simplify the numerical analysis. Consequently, the observational bound translates directly into a constraint on the coupling constant, yielding $\xi < 10^{-11} b^{-2}$. Moreover, using the expression for the massive mode of the Bumblebee field given in Eq.~\eqref{eq11}, the potential coupling strength is expressed as
\begin{equation}
\lambda = \frac{m_B^{2} c^{2}}{2 \hbar^{2} b^{2}} \approx 8.35 \times 10^{7}\,\text{J}^{-1}\,\text{s}^{-1}.
\end{equation}

With all these definitions established, the limiting behavior of the equations of motion \eqref{eq15} and \eqref{eq09} can now be examined in the regime where $B^{2} \to 0$ while $B \dot{B}$ remains nonvanishing. In this limit, one obtains
\begin{equation}
    \frac{\ddot{a}}{a}=\frac{2\lambda k c^2 b^2}{3\xi}\equiv H^2\label{eq13}
\end{equation}
and
\begin{equation}
B\dot{B}=\frac{1}{\xi H}\left[H^2-\frac{\xi H b^2}{4}-H_0^2\right].
\end{equation}

If this situation is interpreted as corresponding to the early universe, namely the inflationary era, Eq.~\eqref{eq13} can be evaluated by assuming that the Hubble parameter saturates the observational upper bound given in Eq.~\eqref{eq14}. This assumption leads to the estimate $\xi \approx 2.9 \times 10^{-87}$ J$^{-1}$ s$^{-1}$ m$^{2}$, which serves as a reference value for the coupling magnitude at this scale. In this regime, the square Bumblebee field $B^2$, originating from zero, is characterized by a brief initial evolution with a variation rate on the order of $\sim 10^{125}$ J s$^2$ m$^{-2}$. This would imply a pronounced exponential expansion of the universe, accompanied by a rapid and significant variation in the intensity of the Bumblebee field.  In other words, using this set of parameters, we can establish an upper bound for the universe's initial era, leading to an estimate of its dynamics.

After this stage, according to Eqs.~\eqref{eq15} and \eqref{eq09}, the dynamics can be expressed as
\begin{equation}
\frac{\ddot{a}}{a}\approx-\frac{2\lambda kc^2B^2}{3\xi},\label{eq16}
\end{equation}
with
\begin{equation}
B\dot{B}\approx\frac{H(1-\xi B^2)}{\xi}-\frac{\lambda k c^2B^4+6H_0^2}{6\xi H}.
\end{equation}
This regime admits a clear physical interpretation. Following the extremely intense exponential expansion, the $B$ field enters the stage described by Eq.~\eqref{eq16} with a large initial amplitude, driving the universe into a phase of abrupt deceleration. Simultaneously, the strong dominance of the $B^4/(\xi H)$ term forces $\dot{B}$ to take large negative values, leading to a rapid suppression of the Bumblebee field intensity.

As time progresses, the influence of the vacuum expectation value (VEV) $b$ becomes increasingly significant, and the dynamics evolve to the complete form,
\begin{equation}
\frac{\ddot{a}}{a}=-\frac{2\lambda kc^2}{3\xi}(B^2-b^2),
\end{equation}
and
\begin{equation}
B\dot{B}=\frac{H(1-\xi B^2)}{\xi}-\frac{\lambda kc^2}{6\xi H}(B^2-b^2)^2-\frac{H_0^2}{\xi H}.\label{eq17}
\end{equation}
In this regime, the value of $B$ gradually decreases, approaching the minimum of the potential, while $H$ exhibits a similar decay behavior.

Assuming that $B^2 \approx b^2$ holds for some interval of time, the equations of motion \eqref{eq08} and \eqref{eq09} reduce to
\begin{equation}
\frac{\ddot{a}}{a}\approx0,\quad\quad B\dot{B}\approx\frac{1}{\xi H}\left(H^2-H_0^2\right).
\end{equation}
During this transient phase, the Bumblebee field oscillates smoothly around its vacuum value $b$, while the cosmic expansion, characterized by $H > H_0$, becomes approximately linear -- indicating a non-accelerated regime. As the Hubble parameter evolves, the field $B$ is driven to slightly higher amplitudes, reinitiating its oscillatory motion with progressively diminishing magnitude.

To better illustrate the underlying dynamics, Eq.~\eqref{eq17} can be rewritten as follows
\begin{align}
\frac{1}{2}\frac{d^2(B^2)}{dt^2}&=-\left[H+\frac{\lambda k c^2(B^2-b^2)}{3\xi H}\right]\frac{d(B^2)}{dt}\nonumber\\&-\left[\dot{H}B^2-\frac{\lambda k c^2(B^2-b^2)^2\dot{H}}{6\xi H^2}\right]\nonumber\\&+\frac{1}{\xi}\left(1+\frac{H_0^2}{H^2}\right)\dot{H},\label{eq22}
\end{align}
which behaves as a forced, damped, harmonic-like oscillator for $B^2 \sim B_\text{eq}^2$, decaying toward an equilibrium configuration. This equilibrium value can be determined by setting Eq.~\eqref{eq22} to zero, yielding
\begin{align}
B^2_\text{eq}&=b^2-\frac{1}{\lambda kc^2}\Biggl\{3H_\text{eq}^2\xi^2+\Biggl[9H_\text{eq}^4\xi^2\nonumber\\&+6\lambda k c^2H_\text{eq}^2(1-b^2\xi)-6\lambda kc^2H_0^2\Biggl]^{\frac{1}{2}}\Biggl\}.\label{eq26}
\end{align}

At the final stage, as the system collapses to the equilibrium configuration characterized by $B^2 = B_\text{eq}^2$ and $H^2 = H_\text{eq}^2$, the condition $\xi H \dot{B} = 0$ is obtained. Considering the relative magnitudes of the constants involved, the relaxation of the system toward equilibrium allows one to take $B_\text{eq}^2 \approx b^2$ and $H_\text{eq} \approx H_0$. Consequently, the dynamics asymptotically return to the regime described by Eq.~\eqref{eq18}, corresponding to the current state of the universe, where $H = H_0$ and the expansion proceeds in an accelerated phase.

In summary, the Bumblebee field is initiated from a singular point -- the origin -- analogously to the early state of the universe. As the $B$ field departs from the equilibrium value $b$, the system finds itself in an unstable configuration, such that even a small (quantum) perturbation is sufficient to trigger a rapid amplification of the field, which subsequently evolves through several intermediate stages. During this process, the Bumblebee field drives both its own variation and the accelerated expansion of the universe toward extremely large negative values, producing a temporary interruption in cosmic evolution and a suppression of Lorentz-violating effects. When $B^2$ approaches the potential minimum $b^2$, the field behaves as a forced, damped harmonic oscillator around the VEV until equilibrium is established, with $H \approx H_0$. To the left of this point, the expansion of the universe proceeds rapidly, whereas to the right, it gradually slows down. Near this stage, the expansion becomes approximately linear, and the evolution of $B$ smoothens. Once equilibrium is reached, the Bumblebee field remains constant, and the cosmic dynamics coincide with those characterizing the present universe, as described by Eq.~\eqref{eq18}.

\subsection{Solutions for \texorpdfstring{$\delta=\frac{3}{2}$}{delta = 3/2}}
\label{secresults3/2}

Here, another case is considered. By setting $\delta = 3/2$, the equations of motion given by Eqs.~\eqref{eq15} and \eqref{eq09} take the form
\begin{equation}
    \frac{\ddot{a}}{a}=-\frac{2\lambda kc^2}{3\xi}(B^2-b^2),\label{eq19}
\end{equation}
and
\begin{equation}
    B\dot{B}=\frac{H(1-\xi B^2)}{\xi}-\frac{\lambda k c^2}{6\xi H}(B^2-b^2)^2-\frac{H_0^2\hbar}{\sqrt{\pi}mc^2\xi}.\label{eq20}
\end{equation}
Here, the onset of the universe proceeds in an almost similar manner: the expansion is extremely rapid, driving the Bumblebee field from zero to very large values within a very short time interval. Subsequently, the expansion of the universe begins to decelerate, and both the Hubble parameter and the Bumblebee field decrease, approaching the potential minimum at $B^2 \approx b^2$. When this regime is attained, the dynamics of the system are governed by the equations
\begin{equation}
    \frac{\ddot{a}}{a}\approx0\quad\quad\text{and}\quad\quad B\dot{B}\approx \frac{H}{\xi}-\frac{H_0^2\hbar}{\sqrt{\pi}mc^2\xi}.\label{eq21}
\end{equation}
This last term in the Bumblebee dynamics, which arises within a non-accelerated expansion regime, differs from the $\delta = 2$ case. It remains constant, shifting the equilibrium point $B^2_\text{eq}$ and rendering the minimum region unstable, thereby inducing a renewed growth of the Lorentz-violating field. The field subsequently decreases toward this point again, causing the cosmic expansion to exhibit a complex, forced, damped, harmonic-like oscillatory behavior that transitions between accelerated, damped, and constant states. As in the $\delta = 2$ case, it can be observed from Eqs.~\eqref{eq19} and \eqref{eq21} that $H$ remains constant when $B^2 < b^2$ and decreases otherwise. According to Eq.~\eqref{eq20}, the Bumblebee field satisfies a forced, damped, harmonic-oscillator-type equation analogous to Eq.~\eqref{eq22}.

The main distinction between the two $\delta$-value cases lies in the position of the equilibrium point. For the present case, according to Eq.~\eqref{eq20}, the equilibrium value of the Bumblebee field is given by
\begin{align}
B^2_\text{eq}&=b^2-\frac{1}{\lambda kc^2}\Biggl\{3H_\text{eq}^2\xi^2+\Biggl[9H_\text{eq}^4\xi^2\nonumber\\&+6\lambda k c^2H_\text{eq}^2(1-b^2\xi)-\frac{6\lambda \hbar kH_0^2}{\sqrt{\pi}m H_\text{eq}}\Biggl]^{\frac{1}{2}}\Biggl\}, 
\end{align}
where the Hubble parameter at equilibrium satisfies
\begin{equation}
\frac{\ddot{a}}{a}=\frac{2\lambda k c^2}{3\xi}(b^2-B^2_\text{eq})\equiv H_\text{eq}^2.\label{eq24}
\end{equation}
Consequently, the field $B$ collapses toward a configuration where $B_\text{eq}^2 < b^2$. This equilibrium lies to the left of that obtained for $\delta = 2$, owing to the presence of the additional term in Eq.~\eqref{eq24}.

After this stage, the Bumblebee field stabilizes at $B^2 = B_\text{eq}^2$, and the cosmic expansion proceeds exponentially, governed solely by the equilibrium Hubble parameter.

\subsection{Solutions for \texorpdfstring{$\delta=1$}{delta = 1}}
\label{secresults1}

For the case where $\delta = 1$, the system of Eqs.~\eqref{eq15} and \eqref{eq09} simplifies to
\begin{equation}
\frac{\ddot{a}}{a}=\frac{2\lambda k c^2}{3\xi}(b^2-B^2),
\end{equation}
together with
\begin{equation}
B\dot{B}=\frac{H(1-\xi B^2)}{\xi}-\frac{\lambda k c^2}{6\xi H}(B^2-b^2)^2-\frac{8\pi^2 mH}{3M_P\xi}.\label{eq25}
\end{equation}
At early times, the qualitative behavior resembles that of the previous cases: the Bumblebee field emerges from an initially vanishing configuration, triggering a phase of rapid accelerated expansion. This process drives $B$ to large values, after which the cosmic expansion gradually slows as the field evolves toward the potential minimum at $B = b$.

A key distinction in this case arises from the fact that, for $\delta = 1$, the last term in the Bumblebee field evolution equation~\eqref{eq25} is directly proportional to the Hubble parameter itself. This dependence causes the field intensity to decrease more rapidly. In contrast, for $\delta = 2$ and $\delta = 3/2$, the corresponding term was proportional to an inverse power of $H$ or remained constant, respectively.

Another fundamental difference emerges when $B^2 \to b^2$, as the system's dynamics is then governed by
\begin{equation}
\frac{\ddot{a}}{a}\approx0, \quad\quad B\dot{B}=\frac{H}{\xi}\left(1-\xi B^2-\frac{8\pi^2 m}{3M_P}\right),
\end{equation}
which, besides leading to a weakened cosmic expansion, induces a renewed increase in the Bumblebee field intensity. Consequently, the evolution of the universe experiences a secondary deceleration phase, driven by the persistence of the nonvanishing $\dot{B}$ term.

On the other hand, $B$ tends to decrease once more, acting as an attractor that drives the field toward an equilibrium configuration characterized by $H = H_\text{eq}$ and $\dot{B} = 0$. The corresponding equilibrium value of the field is given by
\begin{align} 
B_\text{eq}^2&=b^2-\frac{1}{\lambda k c^2}\Biggl\{3H^2_\text{eq}\xi+\Biggl[9H^4_\text{eq}\xi^2\nonumber\\&+6\lambda k c^2H^2_\text{eq}(1-\xi b^2)-\frac{16\lambda kmc^2\pi^2H^2_\text{eq}}{M_P}\Biggl]^{\frac{1}{2}}\Biggr\}
\end{align}
with the Hubble parameter at equilibrium expressed in the same form as Eq.~\eqref{eq24}.

In other words, within the $\delta = 1$ regime, the universe experiences a complex expansion history, alternating between constant, damped, and accelerated phases until the field $B$ settles at the attractor equilibrium point, located slightly away from the VEV $b$, as compared with Eq.~\eqref{eq26}.

 It is worth noting that the dynamic behavior of the Bumblebee field for the three cases analysed here, specifically its evolution towards an equilibrium configuration determined by a symmetry-breaking potential, shares structural similarities with the works in references \cite{referee1, referee2, referee3}. In fact, in those works, there exists a gauge field that spontaneously breaks the Lorentz symmetry, and it is the interplay between the gauge and the Higgs-like fields that drives the accelerated expansion. In this sense, the present model and those references are very similar. However, a distinct feature of the present model is the interplay with the Holographic Dark Energy component. This component then modifies the cosmic dynamics in conjunction with the stabilized vector field, leading to the specific late-time accelerated phase observed.

\begin{figure}[ht]
\includegraphics[scale=0.17]{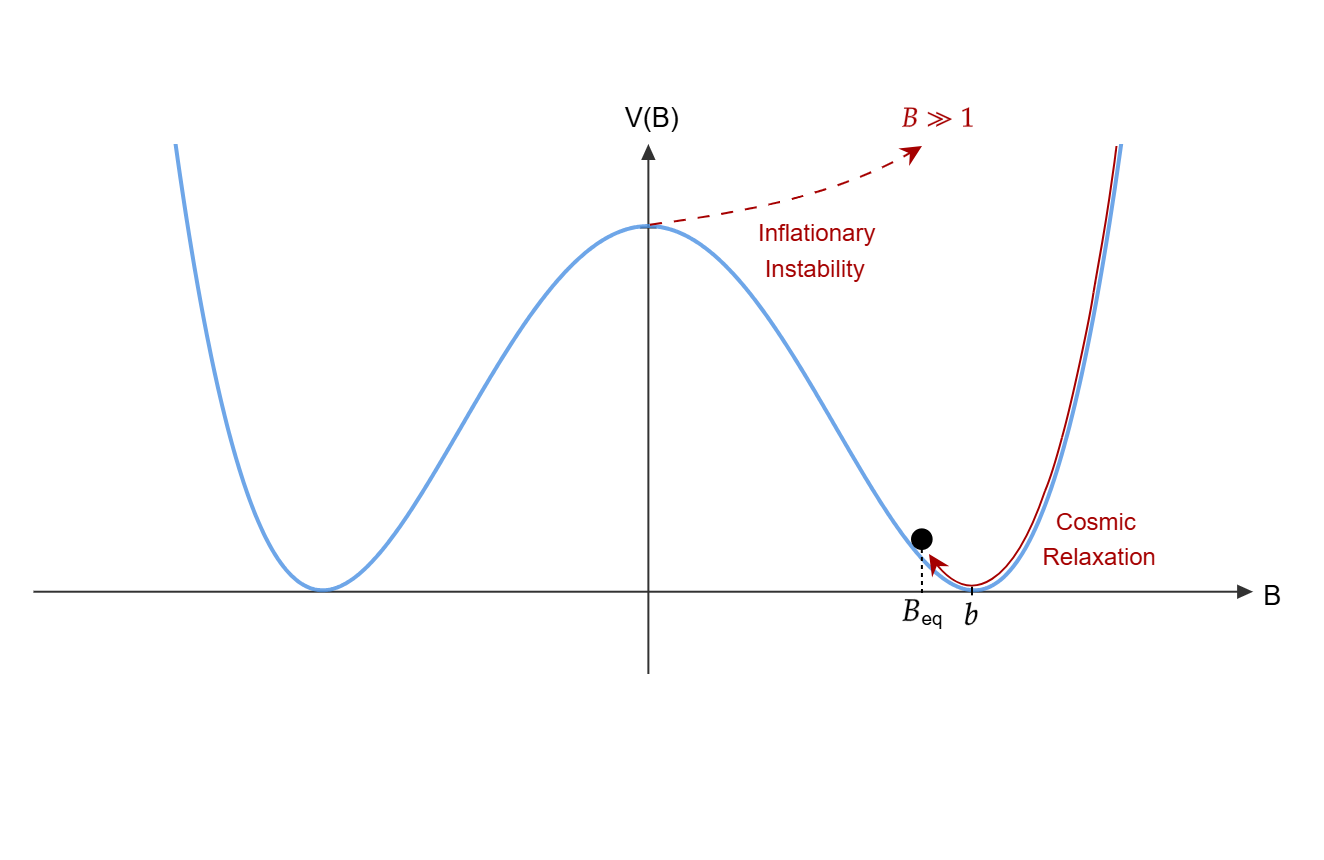}
\caption{Schematic representation of the Bumblebee field dynamics governed by its potential and the interaction with Holographic dark energy.}
\label{fig1}
\end{figure}

 A schematic representation of the dynamics for any chosen $\delta$ values is shown in Figure \ref{fig1}. The diagram illustrates the Bumblebee field evolution, represented by a circle, which initiates at the unstable $B=0$ configuration. The field acquires high values transiently during the early universe before relaxing towards an equilibrium state $B_{eq}$ located near the potential minimum $b$. The displacement between these two points ($B_{eq}$ and $b$) arises from the coupling with the Holographic dark energy component.

Finally, it should be noted that, in the present formulation, only the Tsallis holographic dark energy component has been considered, without the inclusion of ordinary matter. The incorporation of additional cosmic components would render the formulation -- and consequently the discussion -- 
more complete and considerably more intricate. A complete description of the universe and its evolution indeed requires considering all these ingredients to correctly describe the matter and radiation-dominated epochs, ensuring the correct duration for nucleosynthesis and structure formation. However, the present work is focused on the phenomenology of the late-time period, which is dominated by dark energy.

 The final discussion concerning the results addresses their validity and connection with the Hubble parameter issues. This will be discussed in the next section.

 \subsection{Stability, validity and the Hubble tension}\label{secresultsvalidty}

One should note that all solutions for all $\delta$ values describe the complete dynamics of the evolutionary system for each respective regime. The differential equations start from initial values of a vanishing Bumblebee field and the inflation bound of the Hubble parameter given by Eq. \eqref{eq14}, i.e., a limit estimate from early-time measurements \cite{rmeasurement}. This leads the universe’s dynamics to the current state governed by General Relativity, where the quantities are determined by late-time measurements. This provides an alternative point of view on the connection between direct and indirect measurements.

It is essential to clarify the nature of the initial condition chosen for the Hubble parameter in this analysis. The early-time value used here is derived from Planck measurements of CMB anisotropies, which are model-dependent and assume the standard $\Lambda$CDM cosmology. Therefore, the trajectory presented here should be interpreted as a proof-of-concept. We adopt the standard $\Lambda$CDM-inferred value as a base reference to demonstrate that, starting from this established early-universe initial point, the Bumblebee dynamics possess the necessary degrees of freedom to drive the expansion rate towards the local measurements at late times, thereby relieving the tension dynamically.

In addition, the choice of the Bumblebee field described in Eq. \eqref{eq27} might, at first glance, due to the nature of the temporal polarization, give rise to some vacuum instabilities or ghosts. These peculiarities can, in fact, occur in the inflationary regime, where the Bumblebee field, due to fluctuations in the unstable false vacuum state, is driven to very large magnitudes where $B \gg 1$, leading the Lorentz violation to play a primordial dominant role. There, in that strong coupling regime, the Bumblebee theory reaches the limit of its perturbative validity, where such pathologies can emerge. Thus, in that era, namely the initial era, the presented approach is interpreted as an effective theory which needs a fundamental completion to fix the degrees of freedom and issues at high energies.

It is crucial to emphasize that the dynamical behavior connecting early and late-time constraints relies heavily not only on the specific vector field configuration, but also on the chosen coupling constants of Eq. \eqref{eq:LagrangianaGeralBumblebee}. An unaware choice of this set of free parameters can constitute a source of potential pathologies, such as ghosts or Laplacian instabilities. In the context of effective field theories, such pathological modes are problematic if their effects persist within the theory's regime of validity. Thus, while the present setup can successfully addresses the Hubble tension phenomenologically, it represents a theoretical limitation of the current formulation when compared to another fully stable frameworks, suggesting that a UV-complete realization might require a more generalized action.

In other words, a unified description of the cosmological dynamics is provided by Bumblebee gravity with holographic dark energy. Although the inflationary initial era of the universe is described by a background with huge Lorentz violation effects, where the complete theory can be limited by vacuum instabilities, the evolution of the universe acts as a natural relaxing mechanism. This leads the Bumblebee field to a true stable state as an attractor, excluding the Lorentz violation effects and ensuring that the very weak profile of the Bumblebee interaction ($\xi B^2 \approx 0$) describes our present universe free of pathologies. In this framework, the approach presented here gives an alternative formulation addressing the Hubble Tension problem, connecting early- and late-time measurements and fixing the dynamics of $H(t)$ without violating physical restrictions.

\begin{figure}
\includegraphics[scale=0.3]{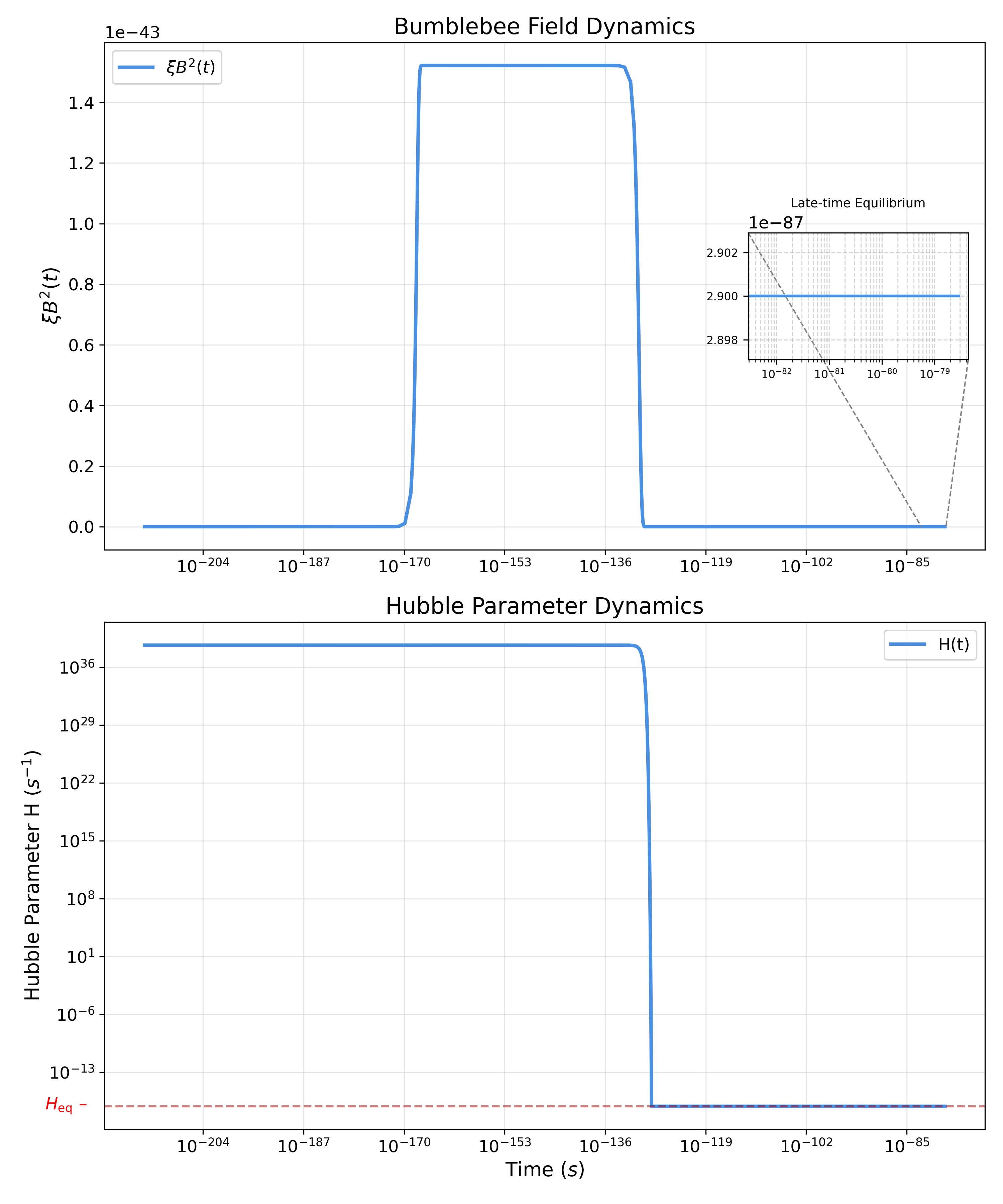}
\caption{Dynamics of the Bumblebee field (normalized by the coupling constant $\xi$) and the Hubble parameter during the first seconds of cosmic evolution for the $\delta=2$ case. The difference obtained here is $b-B\sim10^{-97}$.}
\label{fig2}
\end{figure}

All these features are illustrated in Figure \ref{fig2}, which presents the numerical solutions for the Bumblebee field and Hubble parameter dynamics during the first seconds of cosmic evolution for $\delta=2$ situation. The plot depicts the initial rapid growth of the $B(t)$ field to high values, followed by its decay towards its minimum as $H$ relaxes to the equilibrium point $H_\text{eq}$

\section{Conclusion}
\label{secconclusion}

In this work, the cosmic evolution of a universe governed by the Lorentz-violating Bumblebee framework in the presence of Tsallis holographic dark energy was investigated. The Bumblebee Lagrangian was compared with the gravitational sector of the Standard-Model Extension, allowing bounds on its parameters to be established, and the corresponding equations of motion for the dynamical fields were derived. The behavior of the fields was analyzed in connection with the expansion history of the universe, considering the interaction between the Bumblebee field and dark energy in three distinct holographic formulations. Expressions for the Bumblebee mass and for the dark-energy density were obtained.

The Bumblebee field and the Hubble parameter were shown to be strongly coupled, providing a mechanism in which the initial evolution of the universe emerges from an unstable equilibrium of the Bumblebee potential. This instability is triggered by the non-vanishing vacuum expectation value of the field, which spontaneously breaks Lorentz symmetry by selecting a preferred direction in spacetime, taken to be time-like throughout this work.

 The symmetry breaking occurs through a spontaneous self-interaction mechanism analogous to the Higgs process, introducing a massive sector for the Bumblebee field and triggering a pronounced exponential inflation accompanied by significant Lorentz-violating effects. Following this stage, the Bumblebee dynamics evolve as a damped, forced, oscillator-like system around an equilibrium point which, depending on the holographic formulation, may coincide with the vacuum expectation value $b$. The Hubble parameter, directly coupled to the Bumblebee field, exhibits distinct behaviors across different cosmic eras, including increasing, constant, and rolling-down regimes.  This complete formulation proceeds directly from initial conditions defined by bounds at the inflationary era. It represents an effective theory describing the dynamics of the presented modified gravity after this initial regime, once the Bumblebee field relaxes and Lorentz-violation effects become negligible. This provides an alternative framework to connect early- and late-time measurements, addressing the Hubble tension.

All relevant parameters were presented and compared with observational values, allowing estimates of the symmetry-breaking scale and other quantities that remain experimentally inaccessible. Among these estimates are the gravity–Bumblebee coupling $\xi = 2.9\times10^{-87}\,\text{J}^{-1}\,\text{s}^{-1}\,\text{m}^2$, the Bumblebee mass $m = 2.55\times10^{-3}\,\text{eV}$, and the potential parameter $\lambda = 8.35\times10^{7}\,\text{J}^{-1}\,\text{s}^{-1}$.

Finally, the formalism developed here offers a framework capable of addressing the Hubble tension, providing an alternative means of connecting direct and indirect determinations of the Hubble parameter by relating its inflationary-era value to its present magnitude.

\begin{acknowledgments}

This work by A. F. S. is partially supported by National Council for Scientific and Technological
Development - CNPq project No. 312406/2023-1. D. S. C. and E. M. S. acknowledge financial support from CAPES.
\end{acknowledgments}

\section*{Data Availability Statement}

No Data associated in the manuscript.

\global\long\def\link#1#2{\href{http://eudml.org/#1}{#2}}
 \global\long\def\doi#1#2{\href{http://dx.doi.org/#1}{#2}}
 \global\long\def\arXiv#1#2{\href{http://arxiv.org/abs/#1}{arXiv:#1 [#2]}}
 \global\long\def\arXivOld#1{\href{http://arxiv.org/abs/#1}{arXiv:#1}}



\begin{thebibliography}{99}


\bibitem{Dyson_IX_220_1920} F. W. Dyson, A. S. Eddington and C. Davidson, ``{IX}. {A} determination of the deflection of light by the sun's gravitational field, from observations made at the total eclipse of {May} 29, 1919,'' \doi{10.1098/rsta.1920.0009} {Philosophical Transactions of the Royal Society of London. Series A, Containing Papers of a Mathematical or Physical Character \textbf{220}, 291 (1920)}.

\bibitem{Abbott_Observation_116_2016} B. P. Abbott \textit{et al.}, ``Observation of {Gravitational} {Waves} from a {Binary} {Black} {Hole} {Merger},'' \doi{10.1103/PhysRevLett.116.061102} {Physical Review Letters \textbf{116}, 061102 (2016)}.

\bibitem{Abbott_GW170817_119_2017} B. P. Abbott \textit{et al.}, ``{GW170817}: {Observation} of {Gravitational} {Waves} from a {Binary} {Neutron} {Star} {Inspiral},'' \doi{10.1103/PhysRevLett.119.161101} {Physical Review Letters \textbf{119}, 161101 (2017)}.

\bibitem{Abuter_Detection_615_2018} R. Abuter \textit{et al.}, ``Detection of the gravitational redshift in the orbit of the star {S2} near the {Galactic} centre massive black hole,'' \doi{10.1051/0004-6361/201833718} {Astronomy \& Astrophysics \textbf{615}, L15 (2018)}.

\bibitem{Akiyama_First_875_2019} K. Akiyama \textit{et al.}, ``First {M87} {Event} {Horizon} {Telescope} {Results}. {I}. {The} {Shadow} of the {Supermassive} {Black} {Hole},'' \doi{10.3847/2041-8213/ab0ec7} {The Astrophysical Journal Letters \textbf{875}, L1 (2019)}.

\bibitem{Will_LivRevRel_9_2006} C. M. Will, ``The {Confrontation} between {General} {Relativity} and {Experiment},'' \doi{10.12942/lrr-2006-3} {Living Reviews in Relativity \textbf{9}, 3 (2006)}.

\bibitem{Riess_TheAstJour_116_1998} A. G. Riess \textit{et al.}, ``Observational Evidence from Supernovae for an Accelerating Universe and a Cosmological Constant,'' \doi{10.1086/300499} {The Astronomical Journal \textbf{116}, 1009 (1998)}.

\bibitem{Perlmutter_TheAstrPhysJour_517_1999} S. Perlmutter \textit{et al.}, ``Measurements of Omega and Lambda from 42 {High}-{Redshift} {Supernovae},'' \doi{10.1086/307221} {The Astrophysical Journal \textbf{517}, 565 (1999)}.

\bibitem{PhysToday.40.33(1987)} J. H. Schwarz, ``Superstrings,'' \doi{10.1063/1.881089} {Physics Today \textbf{40}, 33 (1987)}.

\bibitem{BOOK-Springer3(1991)} M. Kaku, ``Introduction to {Superstrings},'' \doi{10.1007/978-1-4684-0397-8_1} {in \textit{Strings, {Conformal} {Fields}, and {Topology}: {An} {Introduction}} (Springer US, 1991) p. 3}.

\bibitem{Rovelli_LivRevRel_11_2008} C. Rovelli, ``Loop {Quantum} {Gravity},'' \doi{10.12942/lrr-2008-5} {Living Reviews in Relativity \textbf{11}, 5 (2008)}.

\bibitem{Carlip_IntJouModPhysD_24_2015} S. Carlip, D.-W. Chiou, W.-T. Ni and R. Woodard, ``Quantum gravity: A brief history of ideas and some prospects,'' \doi{10.1142/S0218271815300281} {International Journal of Modern Physics D \textbf{24}, 1530028 (2015)}.

\bibitem{PhysRevD.39.683(1989)} V. A. Kosteleck\'y and S. Samuel, ``Spontaneous breaking of Lorentz symmetry in string theory,'' \doi{10.1103/PhysRevD.39.683} {Phys. Rev. D \textbf{39}, 683 (1989)}.

\bibitem{NucPhysB.359(1991)} V. A. Kosteleck\'y and R. Potting, ``CPT and strings,'' \doi{10.1016/0550-3213(91)90071-5} {Nuclear Physics B \textbf{359}, 545 (1991)}.

\bibitem{nicolis} A. Nicolis, et. al., ``Zoology of condensed matter: framids, ordinary stuff, extra-ordinary stuff,'' \doi{10.1007/JHEP06(2015)155}{ JHEP \textbf{2015}, 155 (2015)}. 

\bibitem{Universe.2.30(2016)} A. Hees, Q. G. Bailey, A. Bourgoin, H. Pihan-Le Bars, C. Guerlin and C. Le Poncin-Lafitte, ``Tests of {Lorentz} {Symmetry} in the {Gravitational} {Sector},'' \doi{10.3390/universe2040030} {Universe \textbf{2}, 30 (2016)}.

 \bibitem{caoetal} Z. Cao et. al., ``Stringent Tests of Lorentz Invariance Violation from LHAASO Observations of GRB 221009A,'' \doi{10.1103/PhysRevLett.133.071501}{Phys. Rev. Lett. \textbf{133}, 071501 (2024)}.

\bibitem{Colladay_Kostelecky_PhysRevD_55_1997} D. Colladay and V. A. Kosteleck\'y, ``$\mathrm{CPT}$ violation and the standard model,'' \doi{10.1103/PhysRevD.55.6760} {Phys. Rev. D \textbf{55}, 6760 (1997)}.

\bibitem{Colladay_Kostelecky_PhysRevD_58_1998} D. Colladay and V. A. Kosteleck\'y, ``Lorentz-violating extension of the standard model,'' \doi{10.1103/PhysRevD.58.116002} {Phys. Rev. D \textbf{58}, 116002 (1998)}.

\bibitem{gtc1} D. S. Cabral, A. F. Santos and F. C. Khanna, ``Violation of Lorentz symmetries and thermal effects in Compton scattering,'' \doi{10.1140/epjp/s13360-023-03707-w}{The European Physical Journal Plus \textbf{138}, 91 (2023)}.

\bibitem{gtc2} D. S. Cabral and A. F. Santos, ``$e^+e^-\to l^+l^-$ scattering at finite temperature in the presence of a classical background magnetic field,'' \doi{10.1140/epjp/s13360-024-04975-w} {The European Physical Journal Plus \textbf{139}, 190 (2024)}.

\bibitem{gtc3} D. S. Cabral, L. A. S. Evangelista, L. H. A. R. Ferreira and A. F. Santos, ``Electron-positron scattering at finite temperature in Podolsky electrodynamics,'' \doi{10.1016/j.physletb.2025.139874}{Physics Letters B, \textbf{869}, 139874 (2025)}

\bibitem{gtc4} F. Ahmed, J. C. R. de Souza and A. F. Santos, ``Axially symmetric solutions in Ricci-inverse modified gravity,'' \doi{10.1140/epjc/s10052-024-13327-y}{The European Physical Journal C \textbf{84}, 968 (2024)}.

\bibitem{gtc5} F. Ahmed, J. C. R. de Souza and A. F. Santos, ``Time-machines construct in $f(\mathcal{R},\mathcal{A},A^{\mu\nu}A_{\mu\nu})$ and $f(\mathcal{R})$ modified gravity theories,'' \doi{10.1088/1475-7516/2024/10/015}{Journal of Cosmology and Astroparticle Physics, \textbf{2024}, 015 (2024)}.

\bibitem{gtc6} F. Ahmed, J. C. R. de Souza and A. F. Santos, ``An example of rotating cosmological universe in modified gravity,'' \doi{10.1140/epjp/s13360-024-05240-w} {The European Physical Journal Plus \textbf{139}, 419 (2024)}.

\bibitem{PhysRevD.40(1989)} V. A. Kosteleck{\'y} and S. Samuel, ``Gravitational phenomenology in higher-dimensional theories and strings,'' \doi{10.1103/PhysRevD.40.1886} {Physical Review D \textbf{40}, 1886 (1989)}.

\bibitem{PhysRevD.69.105009(2004)} V. A. Kostelecky, ``Gravity, Lorentz violation, and the standard model,'' \doi{10.1103/PhysRevD.69.105009} {Phys. Rev. D \textbf{69}, 105009 (2004)}.

\bibitem{Capelo_PhysRevD_91_2015} D. Capelo and J. P\'aramos, ``Cosmological implications of bumblebee vector models,'' \doi{10.1103/PhysRevD.91.104007} {Phys. Rev. D \textbf{91}, 104007 (2015)}.

\bibitem{Santos_ModPhysLettA_30_2015} A. F. Santos, W. D. R. Jesus, J. R. Nascimento and A. Yu. Petrov, ``G\"odel solution in the bumblebee gravity,'' \doi{10.1142/S021773231550011X} {Modern Physics Letters A \textbf{30}, 1550011 (2015)}.

\bibitem{Jesus_ModPhysLettA_34_2019} W. D. R. Jesus and A. F. Santos, ``Ricci dark energy in bumblebee gravity model,'' \doi{10.1142/S0217732319501712} {Modern Physics Letters A \textbf{34}, 1950171 (2019)}.

\bibitem{Jesus_IntJourModPhysA_35_2020} W. D. R. Jesus and A. F. Santos, ``G\"odel-type universes in bumblebee gravity,'' \doi{10.1142/S0217751X20500505} {International Journal of Modern Physics A \textbf{35}, 2050050 (2020)}.

\bibitem{Caldwell_PhysRevLett_80_1998} R. R. Caldwell, R. Dave and P. J. Steinhardt, ``Cosmological {Imprint} of an {Energy} {Component} with {General} {Equation} of {State},'' \doi{10.1103/PhysRevLett.80.1582} {Physical Review Letters \textbf{80}, 1582 (1998)}.

\bibitem{Gao_PhysRevD_79_2009} C. Gao, F. Wu, X. Chen and Y.-G. Shen, ``Holographic dark energy model from {Ricci} scalar curvature,'' \doi{10.1103/PhysRevD.79.043511} {Physical Review D \textbf{79}, 043511 (2009)}.

\bibitem{Felice_LivRevRel_13_2010} A. De Felice and S. Tsujikawa, ``f({R}) {Theories},'' \doi{10.12942/lrr-2010-3} {Living Reviews in Relativity \textbf{13}, 3 (2010)}.

\bibitem{hubbletension}  E. Di Valentino et al, ``In the realm of the Hubble tension--a review of solutions,'' \doi{10.1088/1361-6382/ac086d} {Class. Quantum Grav. \textbf{38}, 153001 (2021)}.


\bibitem{hubblemod1} T. Adi and E. V. Kovetz, ``Can conformally coupled modified gravity solve the Hubble tension?'' \doi{10.1103/PhysRevD.103.023530}{Phys. Rev. D \textbf{103}, 023530 (2021)}.

\bibitem{hubblemod2} I. Banik and H. Zhao, ``From Galactic Bars to the Hubble Tension: Weighing Up the Astrophysical Evidence for Milgromian Gravity,'' \doi{10.3390/sym14071331}{Symmetry \textbf{14(7)}, 1331 (2022)}.

\bibitem{hubblemod3} T. Schiavone, G. Montani and F. Bombacigno, ``f(R) gravity in the Jordan frame as a paradigm for the Hubble tension,'' \doi{10.1093/mnrasl/slad041}{Monthly Notices of the Royal Astronomical Society: Letters \textbf{522}, 1 (2023)}.

\bibitem{Bekenstein.PhysRevD.7(1973)} J. D. Bekenstein, ``Black Holes and Entropy,'' \doi{10.1103/PhysRevD.7.2333} {Phys. Rev. D \textbf{7}, 2333 (1973)}.

\bibitem{Hawking.CommMathPhys.31(1973)} J. M. Bardeen, B. Carter and S. W. Hawking, ``The four laws of black hole mechanics,'' \doi{10.1007/BF01645742} {Communications in Mathematical Physics \textbf{31}, 161 (1973)}.

\bibitem{Hawking.Nature.248(1974)} S. W. Hawking, ``Black hole explosions?,'' \doi{10.1038/248030a0} {Nature \textbf{248}, 30 (1974)}.

\bibitem{tHooft:1993itb} G. 't Hooft, ``{Dimensional Reduction in Quantum Gravity},'' {in \textit{Salamfestschrift: A Collection of Talks from the Conference on Highlights of Particle and Condensed Matter Physics} (1993) p. 284} [arXiv:gr-qc/9310026].

\bibitem{Tsallis.JornStatPhys.52(1988)} C. Tsallis, ``Possible generalization of Boltzmann-Gibbs statistics,'' \doi{10.1007/BF01016429} {Journal of Statistical Physics \textbf{52}, 479 (1988)}.

\bibitem{Tavayef_PhysLettB_781_2018} M. Tavayef, A. Sheykhi, K. Bamba and H. Moradpour, ``Tsallis holographic dark energy,'' \doi{10.1016/j.physletb.2018.04.001} {Phys. Lett. B \textbf{781}, 195 (2018)}.

\bibitem{Tsalli.EurPhysJC.73(2013)} C. Tsallis and L. J. L. Cirto, ``Black hole thermodynamical entropy,'' \doi{10.1140/epjc/s10052-013-2487-6} {Eur. Phys. J. C \textbf{73}, 1 (2013)}.

\bibitem{Tsallis.PhysLettB.861(2025)} C. Tsallis and H. Jeldtoft Jensen, ``Extensive composable entropy for the analysis of cosmological data,'' \doi{10.1016/j.physletb.2024.139238} {Physics Letters B \textbf{861}, 139238 (2025)}.

\bibitem{Wang.PhysRep.696(2017)} S. Wang, Y. Wang and M. Li, ``Holographic dark energy,'' \doi{10.1016/j.physrep.2017.06.003} {Physics Reports \textbf{696}, 1 (2017)}.

\bibitem{PhysRevD.77.065020(2008)} R. Bluhm, S. Fung and V. A. Kosteleck\'y, ``Spontaneous Lorentz and diffeomorphism violation, massive modes, and gravity,'' \doi{10.1103/PhysRevD.77.065020} {Phys. Rev. D \textbf{77}, 065020 (2008)}.

\bibitem{PhyRevD.74.045001(2006)} Q. G. Bailey and V. A. Kosteleck\'y, ``Signals for {Lorentz} violation in post-{Newtonian} gravity,'' \doi{10.1103/PhysRevD.74.045001} {Physical Review D \textbf{74}, 045001 (2006)}.

\bibitem{hubblevalue1} L. Galbany et al., ``An updated measurement of the Hubble constant from near-infrared observations of Type Ia supernovae,'' \doi{10.1051/0004-6361/202244893} {A\& A \textbf{679}, A95 (2023)}.

\bibitem{hubblevalue2} W. L. Freedman, ``Measurements of the Hubble Constant: Tensions in Perspective,'' \doi{10.3847/1538-4357/ac0e95}{The Astrophysical Journal, \textbf{919}, 16 (2021)}.

\bibitem{planck2028} Planck Collaboration et al., ``Planck 2018 results - VI. Cosmological parameters,'' \doi{10.1051/0004-6361/201833910}{A\& A \textbf{641}, A6 (2020)}.

\bibitem{riess} A. G. Riess, ``Cosmic Distances Calibrated to 1\% Precision with Gaia EDR3 Parallaxes and Hubble Space Telescope Photometry of 75 Milky Way Cepheids Confirm Tension with $\Lambda$CDM,'' \doi{10.3847/2041-8213/abdbaf}{The Astrophysical Journal Letters \textbf{908}, L6 (2021)}.

\bibitem{rmeasurement} Planck Collaboration et al., ``Planck 2018 results - X. Constraints on inflation,'' \doi{10.1051/0004-6361/201833887}{A\&A \textbf{641}, A10 (2020)}.

\bibitem{Li_PhysLettB_603_2004} M. Li, ``A model of holographic dark energy,'' \doi{10.1016/j.physletb.2004.10.014} {Physics Letters B \textbf{603}, 1 (2004)}.

\bibitem{Cohen.PhysRevLett.82(1999)} A. G. Cohen, D. B. Kaplan and A. E. Nelson, ``Effective Field Theory, Black Holes, and the Cosmological Constant,'' \doi{10.1103/PhysRevLett.82.4971} {Phys. Rev. Lett. \textbf{82}, 4971 (1999)}.

\bibitem{hawking2} S. W. Hawking, ``Black holes and thermodynamics,'' \doi{10.1103/PhysRevD.13.191}{Phys. Rev. D \textbf{13}, 191 (1976)}.

\bibitem{higgs1} F. Englert and R. Brout, ``Broken Symmetry and the Mass of Gauge Vector Mesons,'' \doi{10.1103/PhysRevLett.13.321}{Phys. Rev. Lett. \textbf{13}, 321 (1964)}

\bibitem{higgs2} P. Higgs, ``Broken Symmetries and the Masses of Gauge Bosons,'' \doi{10.1103/PhysRevLett.13.508}{Phys. Rev. Lett. \textbf{13}, 508 (1964).}

\bibitem{proca} A. Proca, ``Sur la th\'eorie ondulatoire des \'electrons positifs et n\'egatifs,'' \doi{10.1051/jphysrad:0193600708034700} {J. Phys. Radium \textbf{7}, 8 (1936)}.

\bibitem{kosteleckydatatable} V. A. Kosteleck\'y and N. Russel, ``Data tables for Lorentz and $CPT$ violation,'' \doi{10.1103/RevModPhys.83.11}{Rev. Mod. Phys. \textbf{83}, 11 (2011)}.

\bibitem{referee1} M. Rinaldi, “Dark energy as a fixed point of the Einstein Yang-Mills Higgs equations”, \doi{10.1088/1475-7516/2015/10/023}{JCAP \textbf{2015}, 023 (2015)}.

\bibitem{referee2} M. Álvarez et. al., ``Einstein Yang–Mills Higgs dark energy revisited,'' \doi{10.1088/1361-6382/ab3775} {Class. Quant. Grav. \textbf{36}, 195004 (2019)}.

\bibitem{referee3}  J. Bayron Orjuela-Quintana et. al. ``Anisotropic Einstein Yang-Mills Higgs dark energy,'' \doi{10.1088/1475-7516/2020/10/019} {JCAP \textbf{2020}, 019 (2020)}.



\end{thebibliography}
\end{document}